\documentclass[lettersize,journal]{IEEEtran}
\usepackage{amsmath,amsfonts}
\usepackage{algorithmic}
\usepackage{algorithm}
\usepackage{array}
\usepackage[caption=false,font=normalsize,labelfont=sf,textfont=sf]{subfig}
\usepackage{textcomp}
\usepackage{stfloats}
\usepackage{url}
\usepackage{verbatim}
\usepackage{graphicx}
\usepackage{cite}
\usepackage{makecell}
\begin{document}

\title{SwiftKV: An Edge-Oriented Attention Algorithm and Multi-Head Accelerator for Fast, Efficient LLM Decoding}

\author{Junming Zhang, {\it Graduate Student Member,~IEEE}, Qinyan Zhang, Huajun Sun, {\it Member,~IEEE}, Feiyang Gao, Sheng Hu, Rui Nie, Xiangshui Miao, {\it Senior Member,~IEEE}
\thanks{This work was supported by the National Key Research and Development Program of China(2023YFB4402402) and the National Natural Science Foundation of China (62474074).(Junming Zhang and Qinyan Zhang contributed equally to this work. Corresponding author: Huajun Sun.)

Junming Zhang, Qinyan Zhang, Huajun Sun, Feiyang Gao, Sheng Hu, Rui Nie and Xiangshui Miao are with the School of Integrated Circuits, Hubei Key Laboratory of Advanced Memories, Huazhong University of Science and Technology, Wuhan 430074, China.}}




\maketitle

\begin{abstract}
Edge acceleration for large language models is crucial for their widespread application; however, achieving fast attention inference and efficient decoding on resource-constrained edge accelerators remains challenging. This paper presents SwiftKV Attention, a per-token pipelined, low-latency single-pass attention inference algorithm, where every $(k_t, v_t)$ in the KV cache is processed exactly once in a uniform per-token pipeline without score materialization, blockwise softmax, or a second pass, thereby enabling fast execution on edge accelerators with a single hardware set and no resource-intensive parallelism. Furthermore, to address the limited support for multi-head LLM decoding in existing accelerators, we design the SwiftKV-MHA accelerator, which enables high precision attention and low precision GEMV on the same processor array, achieving fast and efficient multi-head parallel decoding. Experimental results show that, on the edge accelerator, the SwiftKV Attention algorithm achieves a $7.16\times$ speedup over native attention and significantly outperforms other attention algorithms. SwiftKV-MHA further reduces attention latency by $13.48\times$; under the same settings, it improves generation speed by $17.4\%$ and increases token efficiency by $1.98\times$ compared with state-of-the-art works.
\end{abstract}


\section{Introduction}

\IEEEPARstart{T}{he} edge acceleration for large language models (LLMs) is of great significance in reducing system latency, preserving data privacy, and enabling real-time intelligent systems, and has made substantial progress in recent years\cite{ref1,ref2,ref3}. However, when accelerating LLMs on edge accelerators, attention computation latency becomes severe and often emerges as the dominant bottleneck during decode stage. In representative systems, attention accounts for \mbox{33.5\%}\cite{ref4} and \mbox{43\%}\cite{ref5} of end-to-end latency, and the problem worsens with longer contexts—rising to \mbox{60.2\%} at a 16K context length—thereby significantly slowing generation\cite{ref6}.

Existing acceleration strategies on edge accelerators mainly focus on sparsity and quantization~\cite{ref7,ref8,ref9}. 
Traditional online softmax optimizes only the softmax, is not tailored to attention ($QK_{\mathrm{cache}}^{\mathsf{T}}$, $PV_{\mathrm{cache}}$), and still incurs substantial memory traffic from repeatedly accessing attention intermediates~\cite{ref10,ref19}.
Building on this, Flash Attention and related blockwise methods further reduce I/O memory accesses via kernel fusion and blockwise, but are primarily designed for GPU training and prefill ~\cite{ref10,ref11}. 
However, during decoding, the blockwise structure cannot be consistently maintained when newly generated tokens exceed the predefined block boundaries, forcing the computation to wait for block and significantly limiting their effectiveness in decoding.

\begin{figure}[t]
    \centering
    \includegraphics[width=0.7\linewidth, height=0.3\linewidth, keepaspectratio=false]{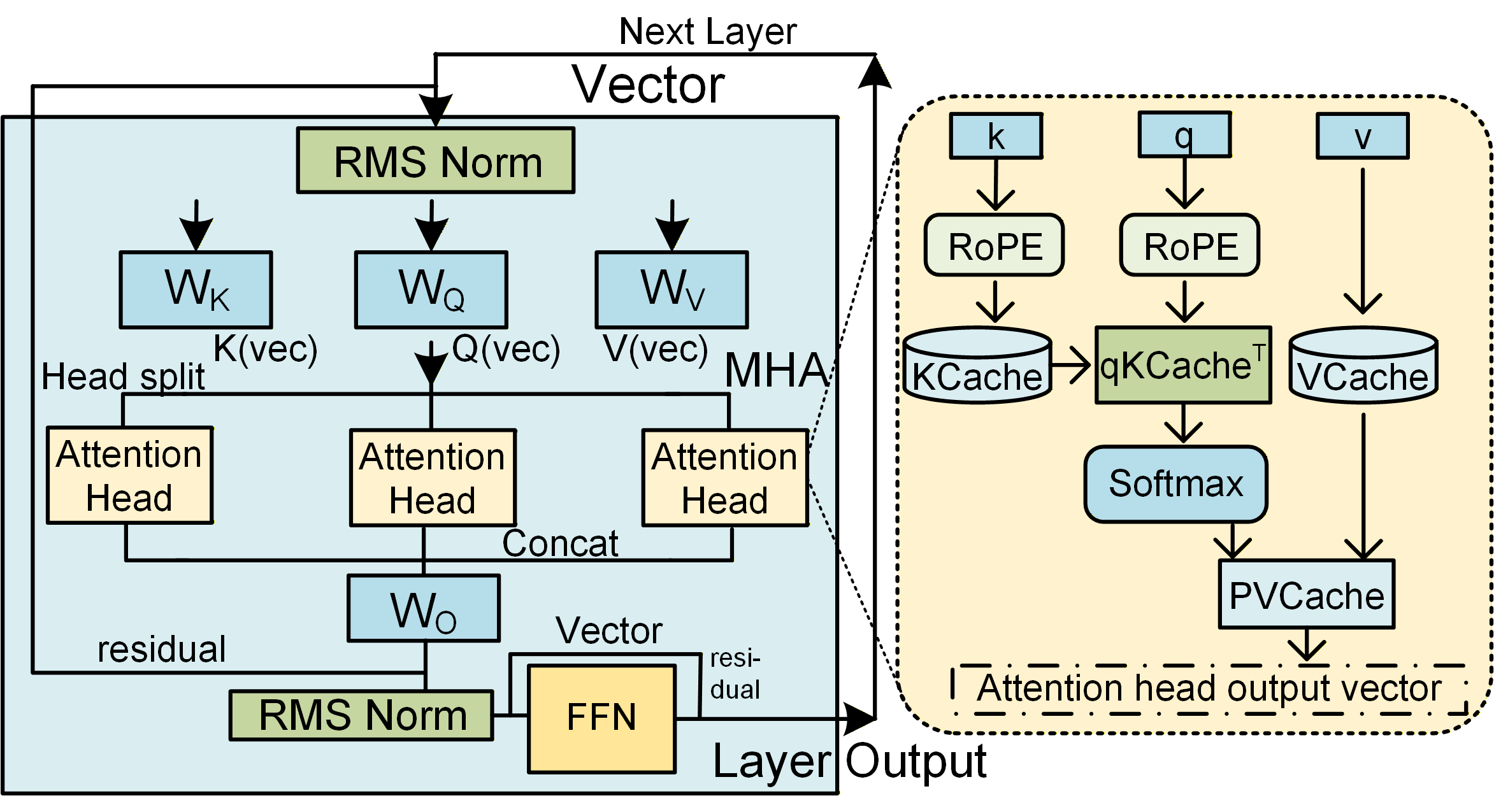}
    \caption{The structure of the multi-head LLM layer during decoding.}
    
\end{figure}

Specifically, on resource-constrained edge accelerators, parallelizing the high-dimensional $\mathbf{q}\mathbf{k}_i^{\mathsf{T}}$, softmax, and subsequent operations is challenging, which slows attention and token generation\cite{ref5,ref12}. As a result, GPU-based blockwise methods become ineffective when only a single set of compute units is available, since block-level parallelism cannot be exploited.
Moreover, fast attention inference often requires dense high-precision multiplications and per-step \textit{exp}\cite{ref10}, leading to a separation from conventional edge accelerators optimized for low-bit GEMV on integer MAC arrays. Therefore, an edge-oriented decode attention algorithm that enables fast, low-latency inference without parallelism and is compatible with low-precision integer computation is urgently needed.

Meanwhile, during decoding in multi-head LLMs, as shown in Fig.1, each head independently executes the full attention (including RoPE). However, conventional GEMM-based acceleration causes frequent cross-module data movement for RoPE, \textit{exp}, softmax operations, and provides insufficient support for multi-head parallelism\cite{ref12,ref13}. 
Currently, there is no processor that can independently complete the entire single-head attention workflow, nor a circuit level RoPE optimized for fast decoding\cite{ref14}, which calls for a decoder-oriented multi-head accelerator.

In this brief, for the edge accelerator, we propose \textbf{SwiftKV Attention}, a fast attention algorithm during decoding, and its corresponding compute engine SwiftKV Core. We further design \textbf{SwiftKV-MHA}, an accelerator optimized for fast multi-head LLM decoding. The contributions are as follows:

\begin{itemize}

  \item We propose \textbf{SwiftKV Attention}, an edge-oriented, per-token pipelined, single-pass attention algorithm that enables low-latency attention on the edge accelerator without additional parallel hardware compute resources.

   \item We design \textbf{SwiftKV-MHA}, a multi-head accelerator with a dual-mode SKV Processor Array that each processor (i) handles per head high-precision  attention inference independently and (ii) performs high throughput low-precision GEMV. It also integrates decoder-specialized RoPE for fast, efficient multi-head LLM decoding.
   
  \item We evaluate and validate the proposed SwiftKV Attention algorithm. We further analyze and compare the performance of the SwiftKV-MHA with state-of-the-art works.

\end{itemize}

\section{Multi-head LLM Decode Structure}

The architecture of one layer in a multi-head LLM during decode is shown in Fig.~1. Each head performs Rotary Positional Embedding (RoPE) on the per-head query ($\mathbf{q}$) and key ($\mathbf{k}$) vectors. 
For position index $m\in\mathbb{N}$ and base $b$ (typically $b=10000$), define angular frequencies:
\begin{equation}
\omega_i = b^{-2(i-1)/d}, \quad i=1,\dots,d/2,
\end{equation}
and rotation angles:
\begin{equation}
\theta_i(m) = m\,\omega_i, \quad i=1,\dots,d/2.
\end{equation}
RoPE rotates each consecutive channel pair by angle $\theta_i(m)$. In matrix--vector form:
\begin{equation}\resizebox{.95\linewidth}{!}{$
\mathrm{RoPE}(\mathbf{q},m)=
\left(
\begin{array}{cccccc}
\cos(m\theta_{0}) & -\sin(m\theta_{0}) & 0 & \cdots & 0 \\
\sin(m\theta_{0}) & \cos(m\theta_{0}) & 0 & \cdots & 0 \\
0 & 0 & \cos(m\theta_{1}) & -\sin(m\theta_{1}) & \cdots \\
0 & 0 & \sin(m\theta_{1}) & \cos(m\theta_{1}) & \cdots \\
\vdots & \vdots & \vdots & \vdots & \ddots \\
0 & 0 & 0 & 0 & \cdots & \cos(m\theta_{d/2-1}) \\
0 & 0 & 0 & 0 & \cdots & \sin(m\theta_{d/2-1})
\end{array}
\right)
\left(
\begin{array}{c}
q_{0} \\ q_{1} \\ q_{2} \\ q_{3} \\ \vdots \\ q_{d-2} \\ q_{d-1}
\end{array}
\right).$}
\end{equation}
The same rotation is applied to $\mathbf{k}$. Attention is then computed:
\begin{equation}
\mathrm{Attention}(\mathbf{q}, \mathbf{K}_{\text{cache}}, \mathbf{V}_{\text{cache}})
=
\mathrm{softmax}\!\left(
    \frac{\mathbf{q}\,\mathbf{K}_{\text{cache}}^{\top}}{\sqrt{d}}
\right)
\mathbf{V}_{\text{cache}} .
\end{equation}

In this work, we adopt two large language models with different parameter scales, \textit{LLaMA2-7B} and \textit{ChatGLM-6B}, for hardware acceleration and validation.

\section{Proposed SwiftKV-Attention and SwiftKV-Core}

Unlike traditional Flash Attention and other attention acceleration algorithms\cite{ref10,ref11} that target prefill or training on GPUs, SwiftKV enables \textbf{per-token pipelined, low-latency single-pass} attention computation on the edge accelerator.

During decoding, $q \in \mathbb{R}^{1 \times d}$ in per head. For every k and v vector $(k_t, v_t)$ in the KV cache, the attention score at token $t$ is computed as
\begin{equation}
s_t = \frac{q k_t^{\top}}{\sqrt{d}}.
\end{equation}
Instead of materializing attention scores or performing blockwise softmax, SwiftKV processes each $(k_t, v_t)$ exactly once.
The score $s_t$ is compared with the $\mu_{t-1}$ from the last attention iteration.
When $s_t \le \mu_{t-1}$,
\begin{equation}
\begin{cases}
\beta = \exp(s_t - \mu_{t-1}),\quad
Z_t = Z_{t-1} + \beta,\\
Y_t = Y_{t-1} + \beta v_t,\quad
\mu_t = \mu_{t-1}.
\end{cases}
\end{equation}
When $s_t > \mu_{t-1}$,
\begin{equation}
\begin{cases}
\alpha = \exp(\mu_{t-1} - s_t),\quad
Z_t = \alpha Z_{t-1} + 1,\\
Y_t = \alpha Y_{t-1} + v_t,\quad
\mu_t = s_t.
\end{cases}
\end{equation}
$Z_t \in \mathbb{R}$ and $Y_t \in \mathbb{R}^{1 \times d}$. Initially, $\mu_1 = s_1$, $Z_0 = 0$, and $Y_0$ is initialized as the zero vector. By deferring the division step, after all k and v vectors in KV cache have been processed, the final attention output is obtained via a one-time normalization:
\begin{equation}
\mathrm{Attention} = \frac{Y_T}{Z_T}.
\end{equation}

In SwiftKV, the exponential factors $\alpha$ and $\beta$ always lie in $(0,1]$, making them suitable for efficient hardware implementation. The exponential is computed as
\begin{equation}
\exp(s_t - \mu_{t-1}) = 2^{(s_t - \mu_{t-1}) \log_2 e} = 2^{n + f},
\end{equation}
where $n$ denotes the integer part, implemented via bit shifting, and $f$ denotes the fractional part.

During decoding, Eqs.~(5)--(9) enable edge accelerators to compute attention as a per-token pipelined, single-pass procedure that scans the full KV cache exactly once. Each incoming $(k_t, v_t)$ is consumed once to update $( Z_t, Y_t)$, avoiding score materialization, revisiting previous tokens or performing any second pass. The computation follows a uniform per-token operation sequence and is naturally pipelined: while post-processing $qk_{t-1}^T$, the next $k_t$ is fetched for the dot-product unit. Since the high-dimensional $qk_{i}^T$ dominating the critical path, all remaining updates can be scheduled within its latency. 
Consequently, low-latency attention inference can be executed efficiently on a single hardware set without resource-intensive parallelism.

Furthermore, SwiftKV adopts 32-bit fixed-point arithmetic (FXP32, Q15.17) for attention, achieving precision better than $10^{-5}$ , which is sufficient for accurate inference. This fixed-point design can reuse the standard MAC arrays used for low-bit integer GEMV, avoiding a dedicated attention datapath.

\begin{figure}[t]
    \centering
    \includegraphics[width=0.95\linewidth, keepaspectratio=false]{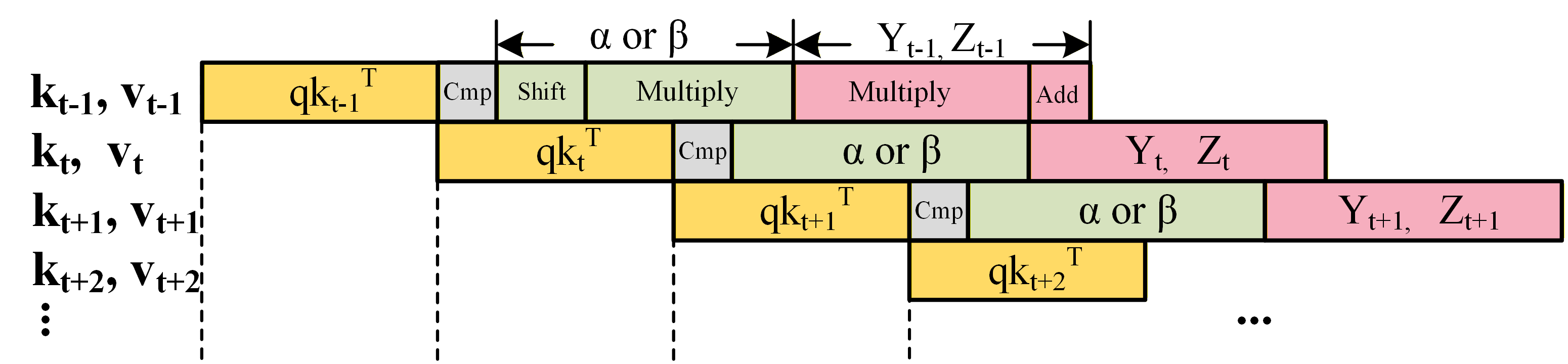}
    \caption{Pipelined computing flow of SwiftKV Attention for edge accelerators.}
    
\end{figure}

\begin{figure}[t]
    \centering
    \includegraphics[width=0.95\linewidth, keepaspectratio=false]{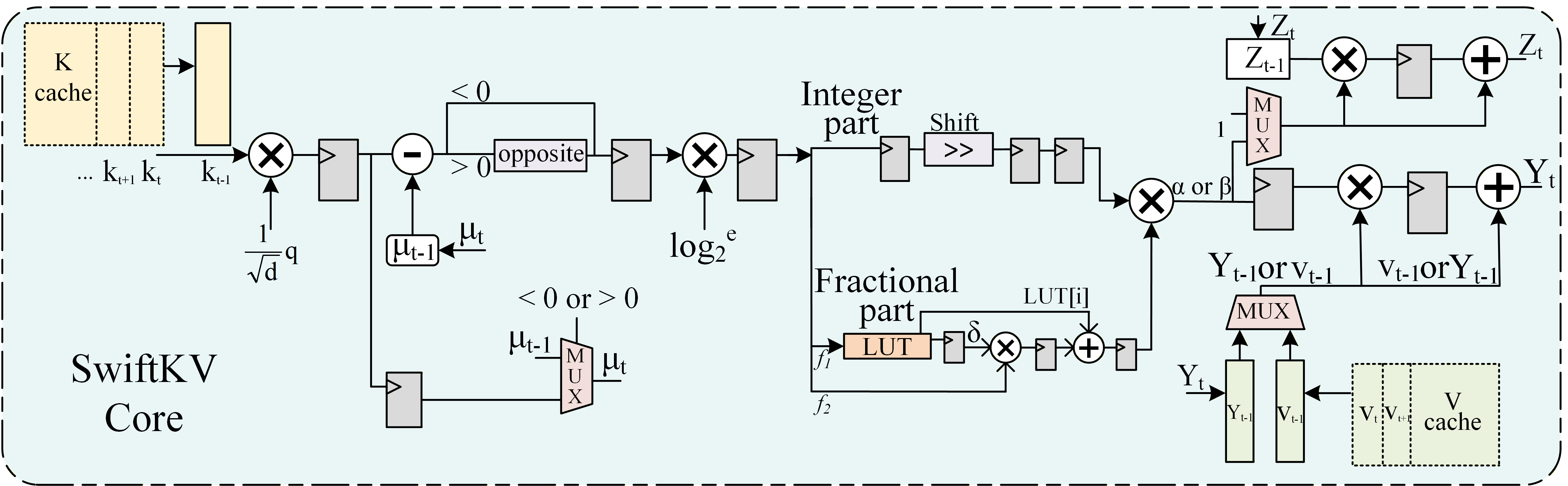}
    \caption{Hardware Compute Engine of SwiftKV-Attention: SwiftKV Core.}
    
\end{figure}

The corresponding hardware acceleration core is illustrated in Fig.~3. The SwiftKV core comprises a single $qk_i^T$ dot-product part, a compare-and-select part, an \textit{exp} part, and an update part for $(Z_t, Y_t)$. For each key--value pair $(k_t, v_t)$ in the KV cache, it supports pipelined KV-cache reads and per-token single-pass attention computation.To compute the fractional term in Eq.~(9) with $f \in (-1,0]$, we approximate $2^{f}$ using a 5-bit lookup table (LUT) with linear interpolation. We decompose $f$ as $f=f_1+f_2$, where $f_1$ contains the 5 most significant fractional bits and $f_2$ contains the remaining 12 bits, yielding an index $i \in \{0,\dots,31\}$. The LUT stores $\mathrm{LUT}[i]=2^{-i/32}$ and corresponding slopes $\delta_i$, and computes
\begin{equation}
2^{f}=\delta_i \cdot f_2 + \mathrm{LUT}[i].
\end{equation}
The overall exponential is then obtained by combining this fractional result with the shift term.

\begin{figure}[t]
    \centering
    \includegraphics[width=0.8\linewidth]{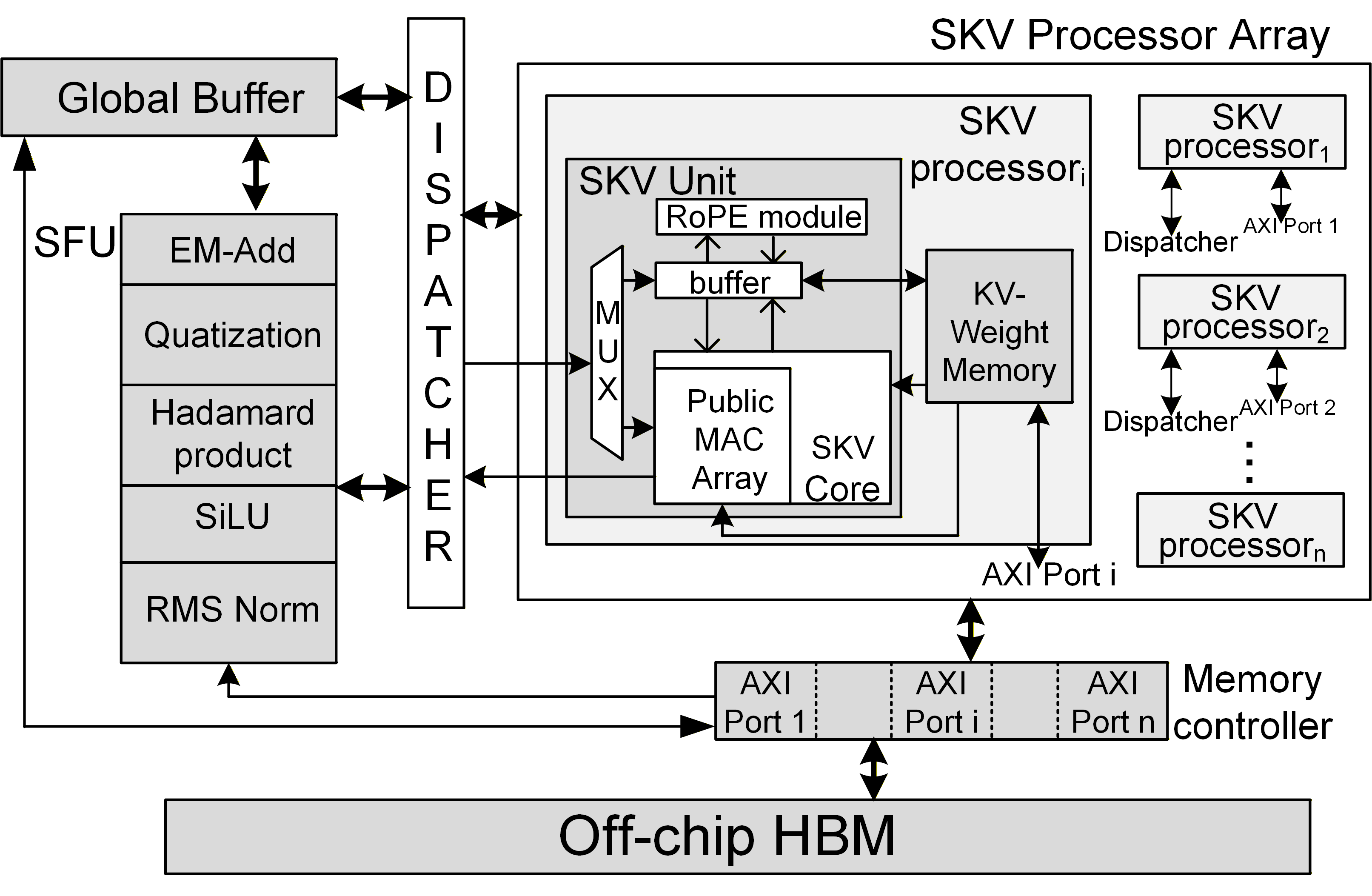}
    \caption{Overall architecture of SwiftKV-MHA accelerator.}
    
\end{figure}

\section{Multi-Head Accelerator Design}
\subsection{Overall Architecture}
The overall architecture of \textbf{SwiftKV-MHA} is shown in Fig.~4. It consists of the SKV Processor Array, Dispatcher, Special Function Unit (SFU), Global Buffer, and Memory Controller. For widely used edge-oriented models in 6B--10B, such as \textit{ChatGLM--6B}, \textit{LLaMA~2--7B}, \textit{LLaMA~3--8B},  \textit{Qwen3--8B}, they mainly adopt a 32-head multi-head attention. The Processor Array includes 32 independent SKV Processors, each handling one head's attention. Each processor has an SKV Unit and a dedicated KV-Weight Memory. The SKV Unit performs RoPE and per-head attention inference independently, while the KV-Weight Memory provides partitioned weight matrices and the KV cache.

The Dispatcher moves data between the Array, Global Buffer, and SFU. It splits high dimensional vectors $x \in \mathbb{R}^{1 \times 4096}$ to 32 processors and collects results into the Global Buffer or forwards them to the SFU. The SFU handles non-MAC operations: elementwise addition (EM-Add), quantization and casting (\texttt{FXP32}/\texttt{INT32}/\texttt{INT8}), Hadamard product, SiLU, and RMS normalization. The Global Buffer stages intermediate vectors ($\mathbf{Q},\mathbf{K},\mathbf{V}$, attention outputs, and FFN activations) and stores each layer’s input and output.

During inference, each Transformer layer operates in \textbf{W4A8} precision. 
The 8-bit input vector $x \in \mathbb{R}^{1 \times 4096}$ is dispatched to the Processor Array to compute $\mathbf{Q}$, $\mathbf{K}$, and $\mathbf{V}$.
For attention, the data are converted in the SFU and split by head; each SKV Processor applies RoPE and computes attention. 
Head outputs are concatenated, converted, and sent back to the Processor Array for FFN computation. 
Finally, the FFN outputs are processed with RMS normalization and SiLU in the SFU and written back to the global buffer for the next layer.

\begin{figure}[t]
    \centering
    \includegraphics[width=\linewidth]{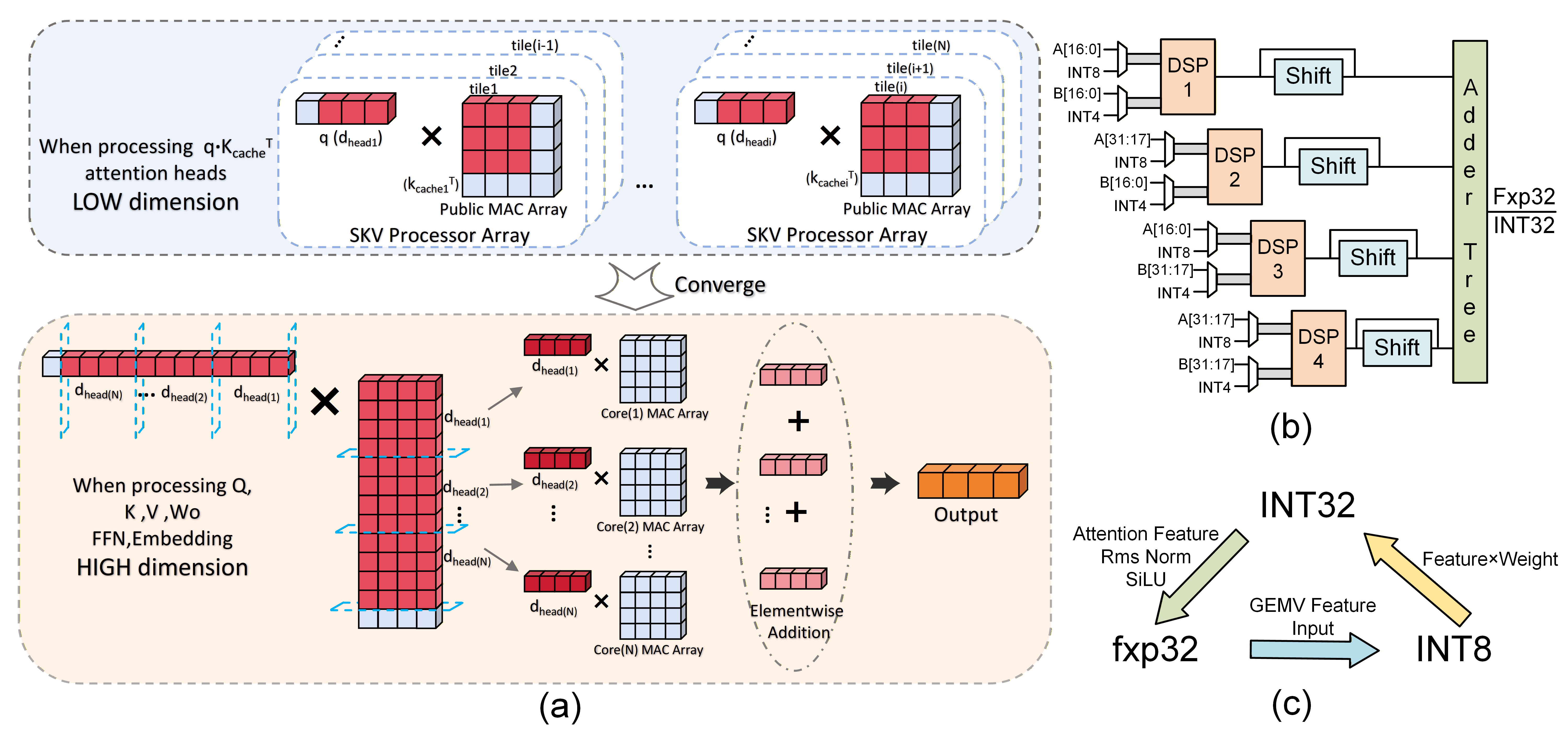}
    \caption{
    Illustration of fast GEMV and attention computation. 
    (a) Using multi-heads to compute per-head attention and high dimensional GEMV. 
    (b) Sharing the same compute resources between high precision FXP32 attention and low precision multi-path GEMV. 
    (c) Precision conversion during computation.
    }
    
\end{figure}

\subsection{Fast GEMV and Attention Computing}
The SKV Processor Array supports both high-precision attention and high throughput low-precision integer GEMV, as shown in Fig.~5(a). When computing attention, for vector $\mathbf{Q} \in \mathbb{R}^{1 \times 4096}$, multi-head attention splits them into 32 heads, each handled by one processor. The Public MAC Array in every SKV Unit computes per head \(q k_i^T\), where \(q,k_i\in \mathbb{R}^{1 \times 128}\).
When executing GEMV between an \texttt{INT8} input and \texttt{INT4} weights, the input is also divided into 32 chunks of 128 dimensions. Each processor multiplies its assigned input chunk with the corresponding weight matrix, producing a partial result, and the 32 partial results are summed using EM-Add in the SFU to form the GEMV output.

The same hardware is shared for both high-precision attention and weight-vector GEMV, enabled by SwiftKV Attention’s use of \texttt{FXP32} arithmetic. 
Both \texttt{FXP32}$\times$\texttt{FXP32}$\rightarrow$\texttt{FXP32} and \texttt{INT4}$\times$\texttt{INT8}$\rightarrow$\texttt{INT32} are treated as fixed-point multiplications (\texttt{FXP(Q15.17)}, \texttt{INT4(Q4.0)}, \texttt{INT8(Q8.0)}), allowing shared use of multiply–accumulate units, as shown in Fig.~5(b).
A typical \texttt{FXP32}$\times$\texttt{FXP32} multiplication requires 4 DSP48E2 units (27×18), while each \texttt{INT4}$\times$\texttt{INT8} uses only one DSP. For GEMV, each Public MAC Array in SKV Unit contains 128 DSPs, enabling a 128-dimensional dot product per cycle. 
With 32 processors operating in parallel, the system completes a 4096-dimensional dot product in a single pass, producing one output element per cycle via pipelining.
For Attention, the format switches to \texttt{FXP32}, where each per-dimension multiplication in the $q \cdot k_i^T$ dot product consumes four DSPs. Thus, the same MAC Array computes a 32-dimensional dot product per cycle, requiring 4 cycles for each $qk_i^T$. With per-token single-pass pipelining (see Fig.~2), Attention over context length $N$ takes about $4N$ cycles. 

As shown in Fig.~5(c), each SKV Processor performs GEMV for $xW_q$, $xW_k$, and $xW_v$, producing \texttt{INT32} partial sums that are quantized to \texttt{FXP32} in the SFU for per-head attention, then back to \texttt{INT8} for $xW_o$. Most data type conversions are overlapped with computation via pipelining, enabling efficient GEMV and attention across layers.

\begin{figure}[t]
    \centering
    \includegraphics[width=0.9\linewidth]{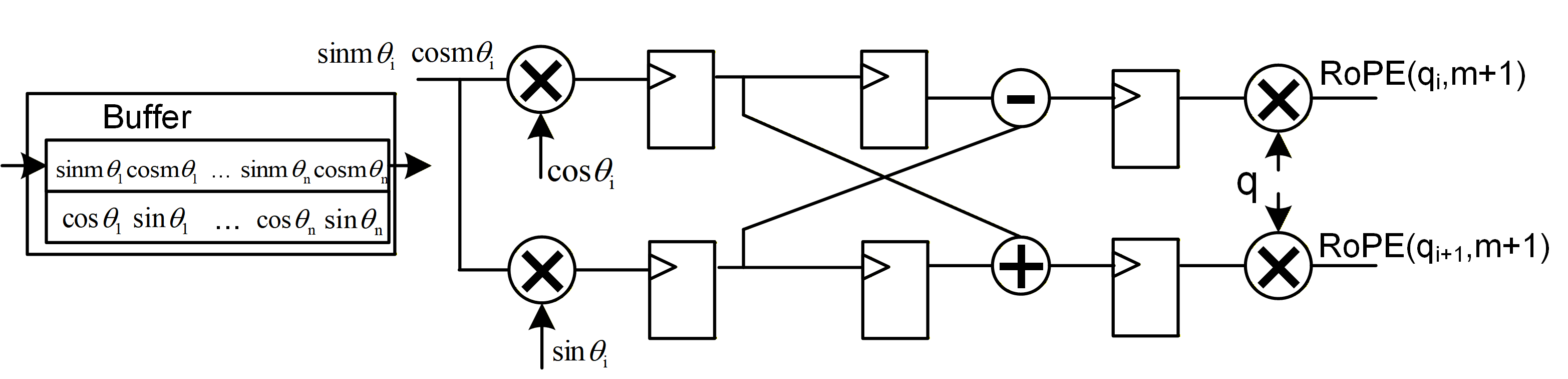}
    \caption{
   The block diagram of decoder-specialized RoPE.}
    
\end{figure}

\subsection{Decoder-Specialized RoPE in Single Head Attention}

Implementing traditional RoPE in hardware is challenging\cite{ref14}: CORDIC only supports angles in $[-\pi/2,\pi/2]$, while RoPE angles $\gamma_{i,j}=i\theta_j$ can be large in LLMs as the context grows. For models such as \textit{LLaMA2-7B} with $\theta_j = 10000^{-j/64}$, directly computing $\cos(i\theta_j)$ and $\sin(i\theta_j)$ is hardware‑expensive.

During decode, tokens are generated sequentially. For a new token at position $m{+}1$, $m$ equals the current context length. We store the initial $\cos(\theta_i)$ and $\sin(\theta_i)$ as constants $a_i$ and $b_i$ in each SKV Unit, $i\in[1,d/2]$, and cache the previous $\cos(m\theta_i)$ and $\sin(m\theta_i)$.
Thus, the RoPE update for the newly generated token becomes:
\begin{equation}
\resizebox{.99\linewidth}{!}{$
\begin{cases}
\text{RoPE}(q_i,m+1)=q_i(a_i\cos m\theta_i-b_i\sin m\theta_i)
-q_{i+1}(a_i\sin m\theta_i+b_i\cos m\theta_i), \\[4pt]
\text{RoPE}(q_{i+1},m+1)=q_i(a_i\sin m\theta_i+b_i\cos m\theta_i)
+q_{i+1}(a_i\cos m\theta_i-b_i\sin m\theta_i).
\end{cases}
$}
\end{equation}

As shown in Fig.~6, this requires only four multipliers, and with pipelining the updated $q_i'$ and $q_{i+1}'$ are produced in three cycles. The new $\text{RoPE}(q,m+1)$ is kept in the unit buffer, while $\text{RoPE}(k,m+1)$ is written to HBM as the updated KV cache. Since all cached keys are already position‑encoded, only the query for the new token requires RoPE, avoiding re-encoding the entire $K$ matrix and significantly reducing decoding complexity.

After $\mathbf{V}$ vector is computed, $\text{RoPE}(q,m+1)$ is read from the buffer and perform attention. As shown in Fig.~4, the full attention head runs independently on all 32 SKV Processors in parallel. Thus, SKV Unit can complete the full single-head attention flow and reduce cross-module data movement and waiting latency.

\section{Experimental Results}

\begin{figure}[htb]
    \centering
    \begin{minipage}[b]{0.49\linewidth}
        \centering
        \centerline{\includegraphics[width=1.6in]{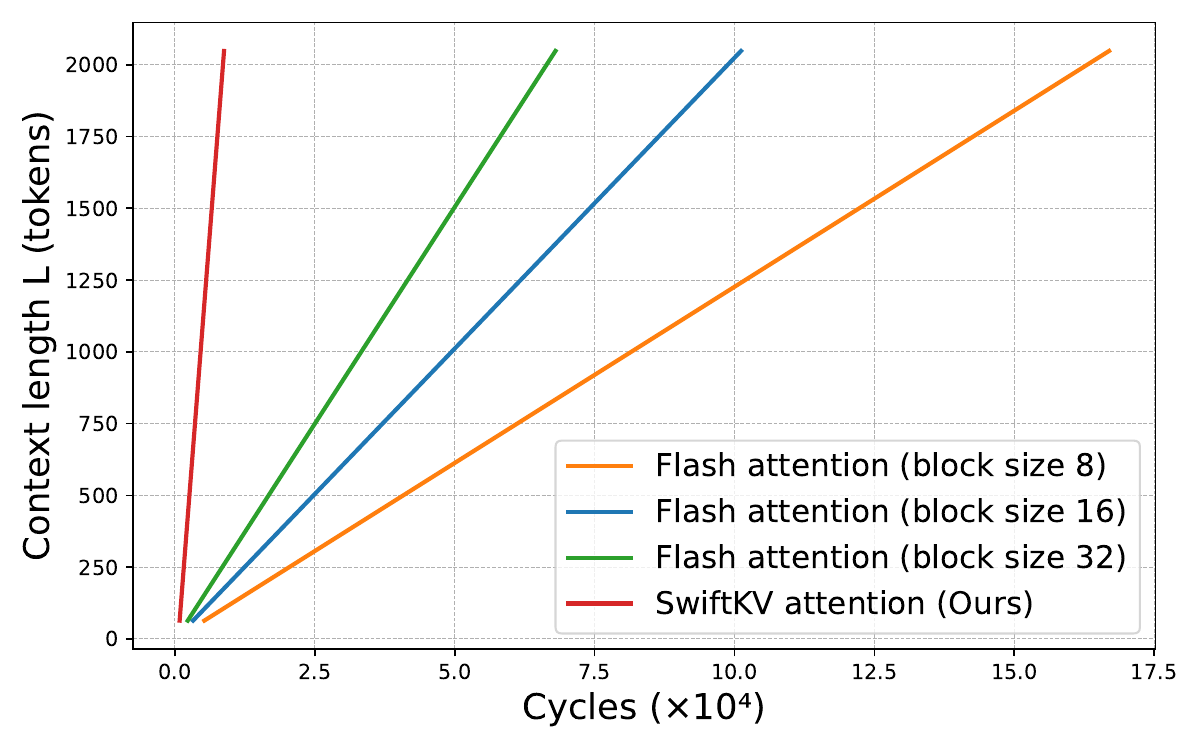}}
    \end{minipage}
    \hfill
    \begin{minipage}[b]{0.49\linewidth}
        \centering
        \centerline{\includegraphics[width=1.74in]{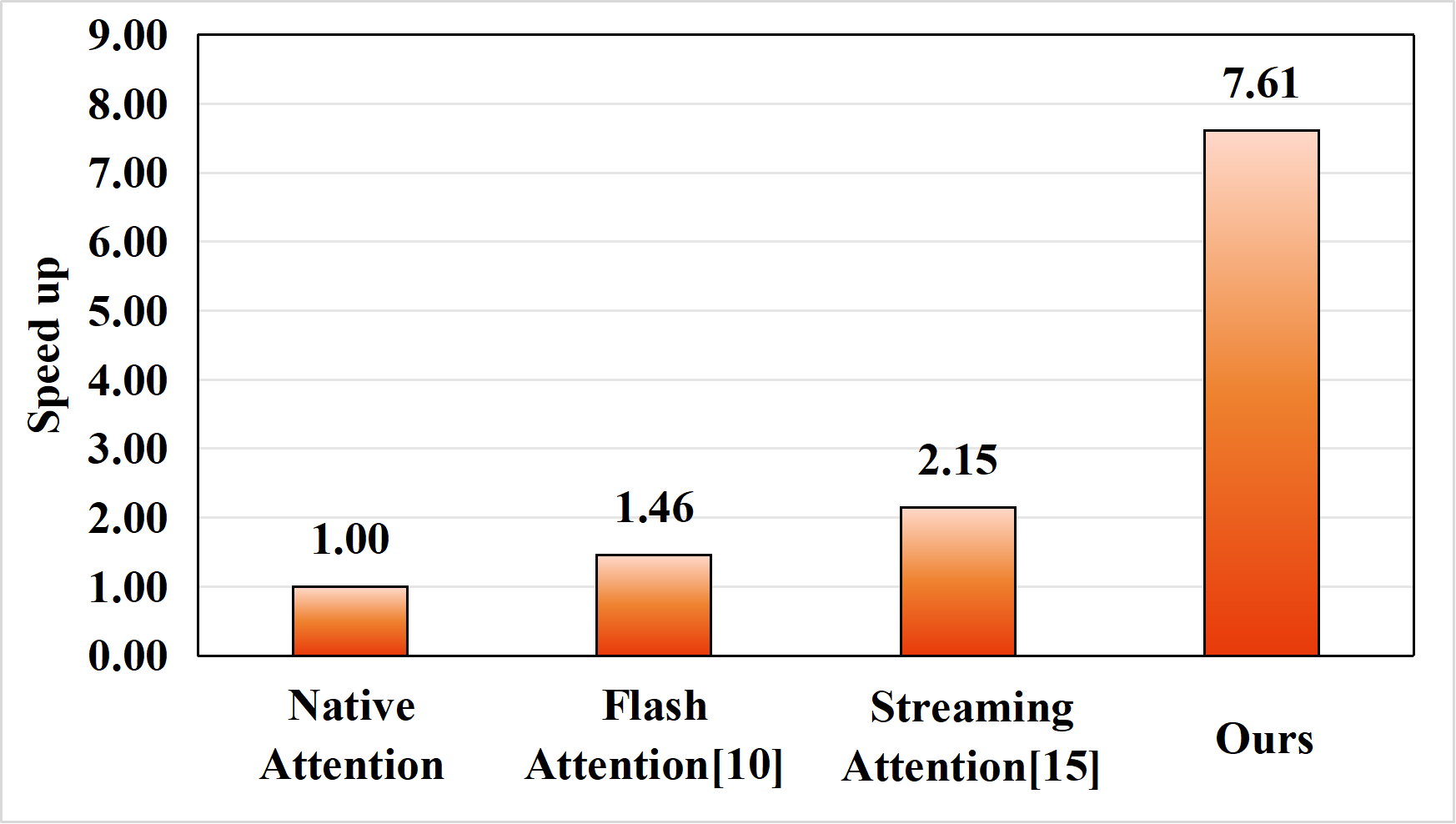}}
    \end{minipage}
    \caption{(a) Attention computation time versus context length for SwiftKV Attention compared with Flash Attention using different block sizes. (b) Speedup comparison of attention computation between SwiftKV Attention and other attention baselines under the same context length.}
\end{figure}

\begin{table}[!t]
\centering
\caption{Token inference accuracy of SwiftKV-MHA}
\begin{tabular}{|c|c|c|c|c|}
\hline
 & \textbf{Top-1} & \textbf{Top-2} & \textbf{Top-3} & \textbf{Top-5} \\
\hline
\textbf{Accuracy} & 100\% & 100\% & 99\% & 98\% \\
\hline
\end{tabular}
\end{table}

\begin{table}[t]
  \caption{Hardware utilization of SwiftKV-MHA on Alveo U55c}
  \label{tab:hardware_util}
  \centering
  \renewcommand{\arraystretch}{1.1}
  \begin{tabular}{lcccc}
    \hline
    \textbf{Component} & \textbf{LUT} & \textbf{FF} & \textbf{BRAM} & \textbf{DSP} \\
    \hline
    \textbf{SFU}             & 14K  & 15K  & 46  & 38 \\
    \textbf{Dispatcher}      & 148K & 65K  & 0   & 0  \\
    \textbf{Processor Array} & 355K & 328K & 224 & 4480 \\
    \textbf{Global Buffer}   & 0    & 0    & 136 & 0 \\
    \hline
    \textbf{Total}  & \makecell{\textbf{517K}\\(39.6\%)} & \makecell{\textbf{408K}\\(15.6\%)} & \makecell{\textbf{406}\\(20.1\%)} & \makecell{\textbf{4518}\\(50.1\%)} \\
    \hline
  \end{tabular}
\end{table}

We evaluate and compare the performance of the SwiftKV attention acceleration algorithm. To ensure a fair comparison, all designs are implemented on the same FPGA platform with the same HBM configuration, using an identical set of \textit{exp} units and the same pipelined multiply and divide units for computing $qK^T$, $PV$, and normalization.

We first compare SwiftKV with traditional Flash Attention using the blockwise approach. As shown in Fig.~7(a), across different context lengths and Flash Attention block sizes (8/16/32), SwiftKV consistently achieves significantly lower attention latency than Flash Attention.
Furthermore, under the same hardware, we compare the attention computation speed of different attention acceleration methods with a fixed context length of 512. As shown in Fig.~7(b), with native attention normalized to 1, Flash Attention(block size 32)~\cite{ref10} achieves a 1.46$\times$ speedup and Streaming Attention~\cite{ref15} achieves a 2.15$\times$, while our SwiftKV reaches 7.16$\times$. This substantially accelerates attention computation for the edge accelerator during decoding and reduces end-to-end latency.

We also evaluate the approximation error of the LUT-based \textit{exp} implementation following Eqs.~(9)-(10). Over the interval $(-1,0]$, the maximum relative error is 0.00586\%, which is sufficient for high-precision forward inference.
To validate accelerator accuracy, we sample 100 sequences of length 512 from the PG-19 dataset and run \textit{LLaMA2-7B}. We sort the output logits and compare the Top-1 to Top-5 tokens against desktop results using the same W4A8 precision. As shown in Table~I, the Top-1 and Top-2 accuracies are both 100\%, indicating its suitability for practical applications. 
The SwiftKV-MHA and experiments are synthesized using Vivado~2022.2 and implemented on an AMD Xilinx U55C board. The resource utilization is reported in Table~II.

The SwiftKV attention algorithm and the SKV processor array capable of both high precision attention computation and high throughput GEMV—the attention latency accounts for only 3.19\% of the total LLM inference latency, as shown in Fig.~8(a). Compared to the original 43.0\% reported in~\cite{ref5}, this represents a 13.48× reduction, confirming the accelerator’s ability to fast compute high precision attention.
For GEMV operations (weight $\times$ activation), the processor array  can perform a dot product of  4096 dimensional vectors in a single cycle. At 225 MHz, this yields a throughput of 1836 GOPS, achieving high performance GEMV computation.

A comparison between SwiftKV-MHA and other accelerators running \textit{LLaMA2-7B} and \textit{ChatGLM-6B} is provided in Table~III. Under the same settings and conditions, the decoding latency per token is only 12.3 ms, resulting in a generation speed of 81.5 tokens per second—17.4\% higher than existing state-of-the-art work\cite{ref9}. The normalized power consumption of the SwiftKV-MHA FPGA chip is 18.3W, with HBM power consumption of approximately 15.5W. The energy efficiency is shown in Fig.~8(b). Compared to prior state-of-the-art works~\cite{ref9,ref13}, the token generation efficiency is improved by 1.98×.
For \textit{LLaMA2-7B}, with a context length of 512, the number of operations required to generate a single token is 13.5 GOP. Thus, the overall throughput of SwiftKV-MHA reaches 1100.3 GOPS (13.5G $\times$ 81.5). Table IV compares our work with different attention-based Transformer model accelerators on FPGA. It is evident that SwiftKV-MHA outperforms previous designs in both throughput and energy efficiency.

\begin{figure}[t]
    \centering
    \includegraphics[width=0.95\linewidth,  keepaspectratio=false]{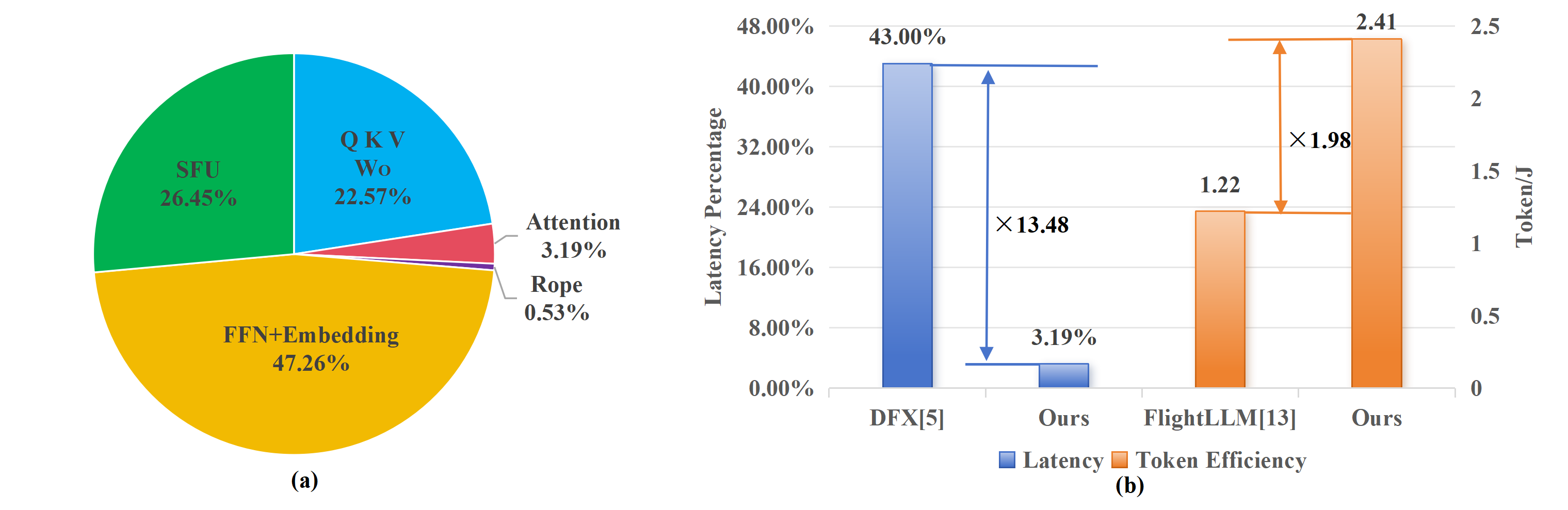}
    \caption{
    (a) Decoding-time latency breakdown of different modules in \textit{Llama2-7B} during inference. (b) Comparison of attention latency (left) and token generation efficiency (right) between SwiftKV-MHA and other accelerators.
    }
\end{figure}

\begin{table}[!t]
\caption{Performance comparison of state-of-the-art FPGA-based Transformer accelerators for LLaMA-2 7B and ChatGLM-6B inference under identical experimental settings}
\centering
\resizebox{\linewidth}{!}{
\begin{tabular}{|c|c|c|c|c|c|}
\hline
\textbf{}  
 & \multicolumn{1}{c|}{\makecell{FlightLLM\cite{ref13} \\ (U280)}} 
 & \multicolumn{2}{c|}{\makecell{EdgeLLM\cite{ref9} \\ (VCU128)}} 
 & \multicolumn{2}{c|}{\textbf{\makecell{This Work \\ (U55C)}}} \\
\hline
\textbf{Model} & Llama-2-7B & Llama-2-7B & ChatGLM-6B & \textbf{Llama-2-7B} & \textbf{ChatGLM-6B}  \\
\hline
\makecell{\textbf{Quantization} \\ \textbf{Type}} & $\sim$W4A8  & \multicolumn{2}{c|}{W4A8} & \multicolumn{2}{c|}{\textbf{W4A8}} \\
\hline
\makecell{\textbf{HBM} \\\textbf{Bandwidth}} & 460\,GB/s & \multicolumn{2}{c|}{460\,GB/s} & \multicolumn{2}{c|}{\textbf{460\,GB/s}} \\
\hline
\makecell{\textbf{Frequency} \\ \textbf{(MHz)}} & 225 & \multicolumn{2}{c|}{225} & \multicolumn{2}{c|}{\textbf{225}} \\
\hline
\textbf{DSP used}  & 6345 & \multicolumn{2}{c|}{4563} & \multicolumn{2}{c|}{\textbf{4518}} \\
\hline
\makecell{\textbf{Latency} \\ \textbf{(ms)}} & 18.2 & 14.4 & 11.7 & \textbf{12.3} & \textbf{10.4} \\
\hline
\makecell{\textbf{Speed} \\ \textbf{(token/s)}} & 55 & 69.4 & 85.8 & \textbf{81.5} & \textbf{96.3} \\
\hline
\makecell{\textbf{System Power} \\ \textbf{(W)}} & 45 & \multicolumn{2}{c|}{56.8} & \multicolumn{2}{c|}{\textbf{33.8}} \\
\hline
\makecell{\textbf{token/J}} & 1.22 & 1.22 & 1.51 & \textbf{2.41} & \textbf{2.85} \\
\hline
\end{tabular}
}
\end{table}

\begin{table}[!t]
\caption{Comparison With Existing FPGA-based works}
\centering
\resizebox{\linewidth}{!}{
\begin{tabular}{|c|c|c|c|c|c|}
\hline
\textbf{Related Works} & \makecell{MICRO'22\\ \cite{ref5}}  & \makecell{TCAS-I'23\\ \cite{ref16}} & \makecell{ASP-DAC'24\\ \cite{ref17}}  & \makecell{TCAS-I'25 \\ \cite{ref18}} & \textbf{\makecell{This Work} }\\
\hline
\textbf{Platform} & Alveo U280  & ZCU102 & Alveo U280 & Alveo U50 & \textbf{Alveo U55C} \\
\hline
\textbf{Model} & GPT2-1.5B & \makecell{Vision\\ Transformer} & BERT-base & \makecell{Swin\\ Transformer} & \textbf{Llama-2-7B} \\
\hline
\makecell{\textbf{Frequency} \\ \textbf{(MHz)}} & 200 & 300 & 220 & 170  & \textbf{225} \\
\hline
\makecell{\textbf{Throughput} \\ \textbf{(GOPS)}} & 184.1  & 726.7 & 757.4 & 830.3 & \textbf{1100.3} \\
\hline
\makecell{\textbf{Efficiency} \\ \textbf{(GOPS/W)}} & 4.09 & 28.2 & 25.1 & 45.12 & \textbf{60.12} \\
\hline
\end{tabular}
}
\end{table}

\section{Conclusion}

We propose SwiftKV Attention, an edge-oriented, per-token, single-pass attention inference algorithm, and SwiftKV-MHA, a fast, efficient multi-head LLM decode accelerator. SwiftKV achieves a $13.48\times$ reduction in attention latency and a $7.16\times$ speedup over native attention, while delivering $17.4\%$ higher speed and $1.98\times$ better efficiency than SOTA works. In future, this solution can be further integrated with AI agents to power practical edge-side LLM applications.

\vfill

\end{document}